\documentclass[epjCONF,columns]{svjour} 

\usepackage[dvips]{graphicx}
\usepackage[varg]{txfonts} 
\usepackage[latin1]{inputenc}

\session-title{The Space Photometry Revolution}

\renewcommand{\l}{\ell}
\newcommand{\eg}{\textit{e.g.}}
\newcommand{\ie}{\textit{i.e.}}
\newcommand{\Teff}{T_{\mathrm{eff}}}
\newcommand{\geff}{g_{\mathrm{eff}}}
\newcommand{\Rpol}{R_{\mathrm{pol}}}

\begin{document}

\title{Internal rapid rotation and its implications for stellar structure and pulsations}
\author{Daniel R. Reese\inst{1}\fnmsep\inst{2}\fnmsep\thanks{\email{dreese@bison.ph.bham.ac.uk}}}

\institute{School of Physics and Astronomy, University of Birmingham, Edgbaston,
Birmingham, B15 2TT, UK \and  Stellar Astrophysics Centre (SAC), Department of
Physics and Astronomy, Aarhus University, Ny Munkegade 120, DK-8000 Aarhus C,
Denmark}

\abstract{Massive and intermediate mass stars play a crucial role in
astrophysics. Indeed, massive stars are the main producers of heavy elements,
explode in supernovae at the end of their short lifetimes, and may be the
progenitors of gamma ray bursts. Intermediate mass stars, although not destined
to explode in supernovae, display similar phenomena, are much more numerous, and
have some of the richest pulsation spectra. A key to understanding these stars
is understanding the effects of rapid rotation on their structure and evolution.
These effects include centrifugal deformation and gravity darkening which can be
observed immediately, and long terms effects such as rotational mixing due to
shear turbulence, which prolong stellar lifetime, modify chemical yields, and
impact the stellar remnant at the end of their lifetime. In order to understand
these effects, a number of models have been and are being developed over the
past few years. These models lead to increasingly sophisticated predictions
which need to be tested through observations. A particularly promising source of
constraints is seismic observations as these may potentially lead to detailed
information on their internal structure. However, before extracting such
information, a number of theoretical and observational hurdles need to be
overcome, not least of which is mode identification. The present proceedings
describe recent progress in modelling these stars and show how an improved
understanding of their pulsations, namely frequency patterns, mode visibilities,
line profile variations, and mode excitation, may help with deciphering seismic
observations.}

\maketitle

\section{Introduction}
\label{intro}

Massive stars play an important role in astrophysics.  Indeed, they are the main
producers of heavy elements, dominate the chemical evolution of galaxies, and
end their lives in supernovae.  Although intermediate mass stars do not explode
in supernovae, they present many of the same phenomena as massive stars, and
furthermore can have very rich pulsation spectra. Therefore, understanding the
structure and evolution of these stars, in particular their chemical evolution
and internal transport processes, is crucial.  Unlike their less massive
counterparts, these stars hardly have any surface convection (apart from
possibly a thin near-surface convective layer). Consequently, they are not spun
down and tend to be rapid rotators (\eg\ \cite{Royer2009}).  Hence, in order to
understand these these stars, one needs to understand the impact of rapid
rotation on stellar structure, chemical evolution, and pulsations.

\section{Impact of rotation on stellar structure and evolution}

Rapid stellar rotation has both immediate effects on stellar structure
as well as long range effects on stellar evolution.

\subsection{Structural changes}

One of the first and most obvious effects of rapid rotation is stellar
deformation.  Indeed, the centrifugal force causes the equator to bulge, as
confirmed by recent interferometric observations (\eg\ 
\cite{DomicianoDeSouza2003}, \cite{Kervella2006}, \cite{DomicianoDeSouza2012}). 
This effect is important because it means these stars can no longer be modelled
using a 1D approach but require a 2D approach.

A second effect, which also has immediate consequences, is gravity darkening. 
Indeed, in rapidly rotating stars, the equator is less luminous than the poles,
due to a smaller vertical temperature gradient.  This effect has also been
observed through interferometry (\eg\ \cite{Peterson2006}, \cite{Monnier2007},
\cite{Che2011}) and is important because it causes the position of these stars
in an HR diagram to depend on their inclination (\eg\ \cite{Suarez2002},
\cite{Salmon2014}).

Various approaches have been used to model this effect, the first being von
Zeipel's law \cite{vonZeipel1924}.  This approach corresponds to the simple
power law $\Teff \propto \geff^{\beta}$, where $\beta = 0.25$.  The value
$\beta=0.8$ has also been proposed for stars with a convective envelope
\cite{Lucy1967}.  More recently, Espinosa Lara and Rieutord came up with a new
law which is based on two assumptions: first, that the luminous flux is parallel
to the effective gravity, and second that the divergence of the luminous flux is
zero (which corresponds to assuming no production of energy in the envelope). 
This leads to an analytical solution which does not depend on any free
parameters, and which compares very favourably with full 2D numerical
simulations from the ESTER code \cite{EspinosaLara2011}.

\subsection{Baroclinic effects}

As opposed to their non-rotating counterparts, rapidly rotating stars have a
baroclinic structure.  In other words, lines of constant pressure, temperature,
and density no longer coincide.  This produces differential rotation which in
turn leads to meridional circulation due to viscosity \cite{Rieutord2005}. 
Currently, only the ESTER code is able to self-consistently calculate such flows
in realistic 2D stellar models (\cite{Rieutord2009}, \cite{EspinosaLara2013}).

As a result of various instabilities, baroclinic flows will then cause
turbulence and enhance transport processes, which modify the stellar lifetime
and chemical yields (\eg\ \cite{Meynet2000}).  Predictions from 1D stellar
evolution codes including a shellular rotation profile based on \cite{Zahn1992}
are in better agreement with various observations, such as the chemical
enrichment of OBA stars, the number ratio of red to blue supergiants etc. (see
\cite{Meynet2005} and references therein).  Recent observations of N enrichment
in O and B stars (\cite{Hunter2008}, \cite{Brott2011b}) have, however, shown
discrepancies with theoretical expectations.  Indeed, there is a lack of
correlation between N enrichment and equatorial velocity, whereas rotational
mixing is needed in theoretical models to produce N enrichment.  It is not clear
what could cause this, although various solutions have been proposed, such as
transport of chemical species by waves \cite{Aerts2014}.

\subsection{Impact on convection zones}

Rapid rotation also affects convection zones.  For instance,
\cite{EspinosaLara2007} produced a rapidly rotating model with a convectively
unstable equatorial belt.  A year later, \cite{Maeder2008}, using a 1D
formalism, found that rotation favours convection in stellar envelopes,
especially at the equator.  Nonetheless, it remains an open question as to what
prescription for convection should be used in a 2D rapidly rotating model, and
is currently the main reason why the ESTER code is unable to model low mass
stars.  However, ongoing work by \cite{Wang} may provide an answer to this
question.  Indeed, they are developing a new convection code based on
unstructured meshes, capable of handling centrifugal deformation. Initial
comparisons with benchmarks from the ASH code show promising results.

\subsection{Summary}

In summary, rotation causes many new phenomena which affect stellar structure,
transport processes, mixing, and evolution.  Although much progress has been
made in our theoretical modelling of these processes, there remains large
uncertainties due to their complexity and to discrepancies with current
observations.  As a result, it is necessary to constrain these processes through
further observations, and asteroseismology is currently the best way of doing
this as it is the only way we have to probe the internal structure of stars. 
However, before being able to carry out asteroseismology, one needs to be able
to model the effects of rapid rotation on stellar pulsations, which, as will be
described in the next section, is by no means trivial.

\section{Impact of rotation on stellar pulsations}

In the non-rotating case and in the absence of phenomena which break spherical
symmetry, such as magnetic fields, stellar pulsation modes with the same radial
order, $n$, and harmonic degree, and $\l$, but differing azimuthal orders, $m$, 
have the same frequencies and are thus degenerate.  However, it has been known
for many years that rotation lifts this degeneracy \cite{Ledoux1951}, in much
the same way that the Zeeman effect splits absorption lines.  The first order
effect of rotation can be split into two terms.  The first is simply an
advection term which leads to a sort of Doppler shift of the frequencies on
non-axisymmetric modes.  The second comes from the effects of the Coriolis force
and is characterised by the Ledoux constant.  Inversions of solar or stellar
rotation profiles rely on this first order approximation (\eg\ \cite{Schou1998},
\cite{Deheuvels2012}, \cite{Kurtz2014}).  At more rapid rotation rates, one
needs to include higher order effects of rotation (\eg\ \cite{Saio1981},
\cite{Soufi1998}), and ultimately apply a 2D approach (\eg\ \cite{Reese2006},
\cite{Ballot2010}).  In what follows, we will describe the effects of rapid
rotation on stellar pulsations, as based on the latest 2D numerical simulations.

\subsection{Gravito-inertial modes}

At the lower end of the frequency spectrum in non-rotating stars, there are
gravity (or g-) modes for which the restoring force is buoyancy.  When the star
is rotating, buoyancy combines with the Coriolis force, thereby leading to
gravito-inertial modes.  The type of stars which display gravito-inertial modes
are  $\gamma$ Dor stars, SPBs, and Be stars.  If the mode owes its existence to
the Coriolis force, then it is known as an inertial mode. An extensive
literature on inertial modes, and singular gravito-inertial modes exist (\eg\
\cite{Papaloizou1978}, \cite{Lee2006}, \cite{Rieutord2000}, \cite{Dintrans2000},
\cite{Mirouh}), but in what follows, we will focus on modes which become
classical g-modes in the non-rotating limit.  

According to Tassoul's asymptotic formula \cite{Tassoul1980}, gravity modes in
non-rotating stars are evenly spaced in period, for a given $\l$ value.  When
the star is rotating, the period spacing also depends on $m$ and on the ratio
$\eta=\frac{2\Omega}{\omega}$ which characterises the effects of the Coriolis
force on the pulsation mode, where $\Omega$ is the rotation rate and $\omega$
the pulsation frequency.  Such a relation was first established using the
traditional approximation (\cite{Berthomieu1978}, \cite{Lee1987}) and has been
subsequently confirmed through full 2D calculations, although the centrifugal
deformation  slightly affects some of the modes \cite{Ballot2012}.

In terms of mode geometry, it is extremely important to distinguish between
sub-inertial modes, for which $\omega < 2 \Omega$, and super-inertial modes,
where $\omega > 2 \Omega$.  Indeed, in the sub-inertial regime, modes are
confined to an equatorial region thanks to critical surfaces, as described in
\cite{Dintrans1999} and \cite{Dintrans2000}.  In the super-inertial regime,
gravito-inertial modes tend to behave like their non-rotating counterparts,
except for one notable exception -- rosette modes. Such modes where discovered
by \cite{Ballot2012} using a 2D numerical approach, and have subsequently been
studied in detail by \cite{Takata2013}, \cite{Saio2014}, \cite{Takata2014},
\cite{Takata} using a variety of techniques.

\subsection{Acoustic modes}

At the upper end of the frequency spectrum, there are acoustic (or p-) modes for
which the restoring force is pressure.  Given that these modes tend to be
located in the outer portions of the star, they are most affected by centrifugal
deformation, which is greatest in such regions.  One way of characterising the
effects of the centrifugal deformation is by taking the ratio of the
characteristic length scale of the deformation, which is roughly proportional to
$\Omega^2$, to the mode's wavelength, which is proportional to $\omega^{-1}$.
Hence, high frequency modes are more affected than low frequency modes, due to
their smaller wavelength.  $\delta$ Scuti and $\beta$ Cephei stars oscillate in
the acoustic domain and typically tend to be rapid rotators.

At rapid rotation rates, the acoustic modes subdivide into different classes of
modes, each with their own characteristic geometry and independent frequency
organisation.  This was first shown in polytropic models by
\cite{Lignieres2008}, \cite{Lignieres2009} using 2D numerical simulations and
ray dynamics and was then extended to more realistic models using 2D numerical
simulations by \cite{Reese2009}.

\begin{figure}[htbp]
\includegraphics[width=\columnwidth]{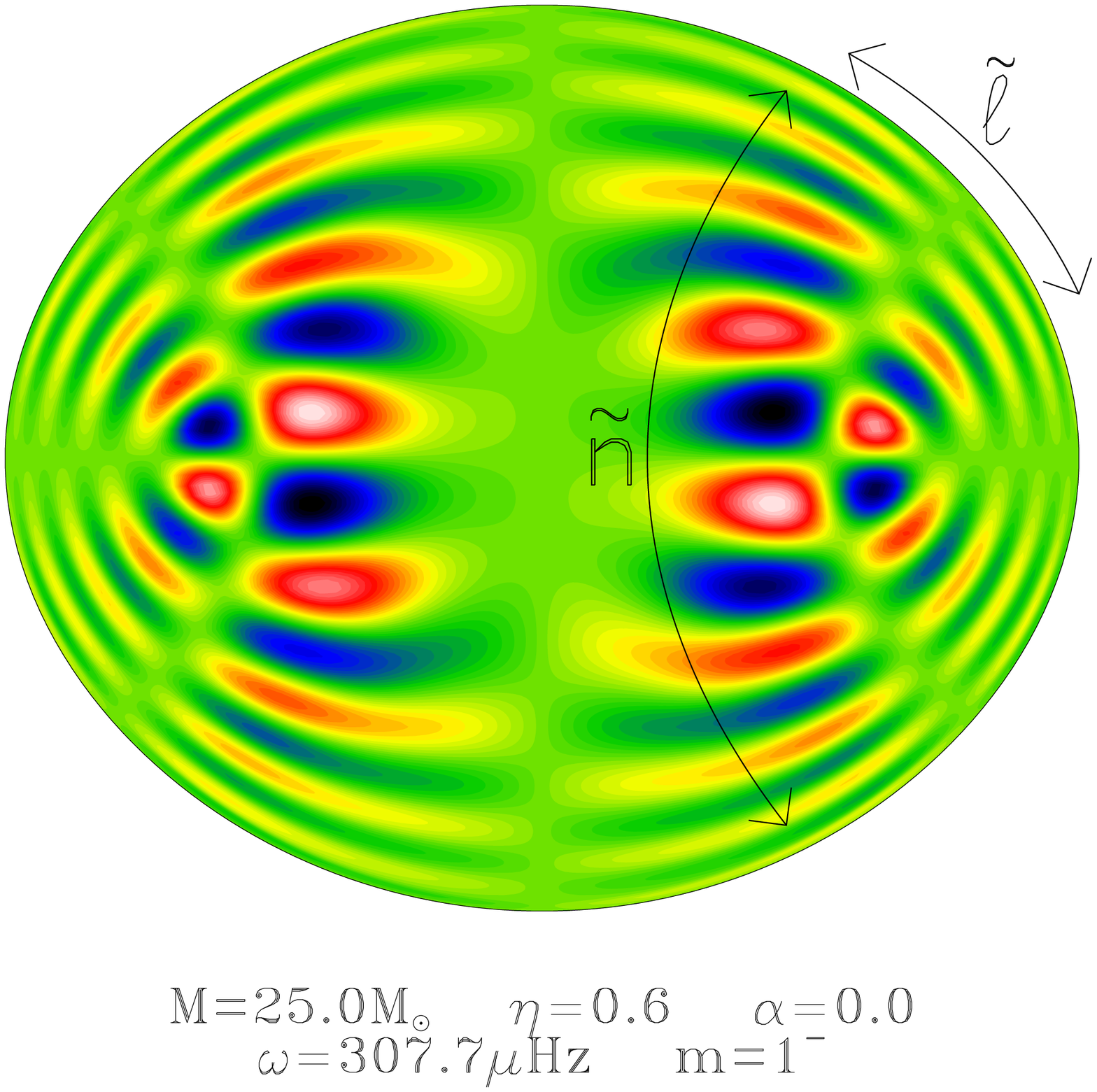}
\caption{Island mode in a model based on the Self-Consistent Field method
\cite{MacGregor2007}.  New quantum numbers are indicated by the
arrows.\label{fig:island}}
\end{figure}

Among these different classes, island modes are the most important, since they
are the most visible of the regular (non-chaotic) modes.  These modes are the
rotating counterpart to modes with few nodes from one pole to another, \ie\ with
a small $\l-|m|$ value.  Their geometric structure closely follows a 2-period
ray orbit and is characterised by new quantum numbers as illustrated in
Fig.~\ref{fig:island}.  Using these quantum numbers, it is possible to obtain a
new asymptotic formula which describes the frequencies of these modes, the
coefficients of which are related to travel time integrals (see
\cite{Lignieres2008}, \cite{Lignieres2009}, \cite{Pasek2011} and
\cite{Pasek2012}).  Of particular interest is the frequency spacing,
$\Delta_{\tilde{n}}$, between modes with consecutive $\tilde{n}$ values.  This
spacing is, in fact, half the large separation, and roughly scales with the
square-root of the stellar mean density, even a rapid rotation rates
\cite{Reese2008}.

\subsection{Mixed modes}

Mixed modes combine the characteristics of acoustic and gravity modes, and occur
when there are avoided crossings between these modes.  Such modes typically
occur in evolved stars, but can also be found in unevolved rapidly rotating
stars.  Indeed, gravito-inertial modes are hardly affected by centrifugal
deformation and hence tend to scale as $\sqrt{GM/\Rpol^3}$, where $\Rpol$ is the
polar radius.  In contrast, acoustic modes and the characteristic spacing
$\Delta_{\tilde{n}}$ are roughly proportional to $\sqrt{GM/V}$, where $V$ is the
volume of the star.  As a result, the frequency domain of acoustic modes
decreases compared with that of gravito-inertial modes, thereby causing the two
to overlap. This favours avoided crossings and hence mixed modes, especially
given that rotation allows coupling between modes with different harmonic
degrees.

In evolved stars, rotation may play an important role in mixed modes.  Indeed,
as shown by \cite{Ouazzani2013}, rotation affects different members of a given
rotation multiplet differently, even at relatively small rotation rates.  This
is because the avoided crossings which produce the mixed modes occur at slightly
different values of the rotation rate for different members of the multiplet. 
As a result, the ratio of the p- and g-mode contributions will be different,
thereby leading to different mode inertias, and the frequencies will no
longer be evenly spaced.

\section{Asteroseismology}

Having described the effects of rapid rotation on stellar pulsations, we now
turn our attention to the asteroseismic inferences which can be drawn from
observed stellar pulsations.  In what follows, we will distinguish between
average or global asteroseismology, which focuses on the general
characteristics of the pulsation spectrum, and detailed or ``boutique''
asteroseismology, which relies on identifying individual modes.

\subsection{Global asteroseismology}

As was done above, we first start with the lower end of the frequency spectrum. 
Recent observations have shown that the pulsation frequencies of
gravito-inertial modes in a number of Be stars tend to clump together.  As
explained in various publications (\eg\  \cite{Walker2005b}, \cite{Saio2007},
\cite{Cameron2008}), the co-rotating frequencies of the pulsation modes are
likely to be much smaller than the rotation rate.  When viewed from an inertial
frame, these frequencies are Doppler shifted by $m\Omega$ as described above,
thereby leading to separate clumps for each azimuthal order. Accordingly, the
frequency differences between these clumps provide an asteroseismic measure of
the rotation rate.  However, a recent publication, \cite{Semaan2013}, showed
discrepancies between this seismic rotation rate and more classical
measurements of $\Omega$ based on spectroscopy.

At higher frequencies, recent studies of p-modes in rapidly rotating $\delta$
Scuti stars have found recurrent frequency spacings using histograms of
frequency differences or Fourier transforms of frequency spectra
(\cite{Breger2009}, \cite{GarciaHernandez2009}, \cite{Mantegazza2012},
\cite{GarciaHernandez2013}, \cite{Suarez2014}, \cite{GarciaHernandez}).  Similar
studies have also been carried out using theoretical pulsation spectra based on
realistic mode visibilities (\cite{Lignieres2010}, \cite{Reese}) and have shown
that in favourable cases, such recurrent spacings could correspond to the large
frequency spacing or half its value (\ie\ $\Delta_{\tilde{n}}$), or to a
multiple of the rotation rate.  Interpreting these spacings as the large
separation, \cite{GarciaHernandez2013} then went on to constrain the mean
density of a $\delta$ Scuti star observed with CoRoT.

\subsection{Detailed asteroseismology}

In order to obtain tighter constraints on stellar properties, one needs to carry
out detailed asteroseismology.  However, up until now, it has proven to be very
difficult to correctly match observed frequencies with theoretical modes, \ie\ 
identify modes, in rapidly rotating stars due to the lack of \textit{simple}
frequency patterns, as is very well illustrated in Fig.~5 of
\cite{Deupree2012}.  In order to overcome these difficulties, one can envisage
various approaches.  One can still try to look for the more complicated patterns
through the use of Echelle diagrams (\cite{GarciaHernandez2013},
\cite{GarciaHernandez}, \cite{Bedding}) or by matching an asymptotic formula to
the frequencies \cite{Reese2009b}.  The latter approach will however be thrown
off by the presence of chaotic modes which follow their own frequency
distribution \cite{Lignieres2009}.  Another possibility is to apply
observational mode identification techniques based on multi-colour photometry or
spectroscopy.  Such techniques have already been applied in slowly rotating
stars, but they need to be adapted to more rapid rotation.

In the photometric approach, one typically looks at the amplitude ratios or
phase differences of a given pulsation mode in different photometric bands.  The
advantage of this approach is that the intrinsic mode amplitude factors out,
thereby leaving information which is only based on mode geometry.   As opposed
to the non-rotating case, the amplitude ratios and phase differences depend both
on the azimuthal order and on the inclination (\eg\ \cite{Townsend2003},
\cite{Daszynska_Daszkiewicz2007}).  However, at a given inclination, similar
amplitude ratios are obtained for modes with the same $\l$ and $m$  but
different radial orders \cite{Reese2013}.  This is due to the similar surface
geometric structure of these modes, as expected from asymptotic theory
\cite{Pasek2012}.  Accordingly, by grouping together modes with similar
amplitude ratios, one can find families of modes with similar quantum numbers,
thereby constraining mode identification and characteristic spacings such as the
large frequency separation or the rotation rate \cite{Reese}.

The spectroscopic approach consists in obtaining a time series of high
resolution spectra of a pulsating star to see how the absorption lines vary over
time as a result of Doppler shifts from the oscillatory motions, and comparing
this to theoretical predictions.  Such an approach cannot currently be carried
out from space and requires the use of ground based-telescopes, preferably in a
network such as SONG.  Previous and current spectroscopic observations of
rapidly rotating pulsating stars include those described in \cite{Telting1998},
\cite{Poretti2009}, \cite{Themessl}.  Various studies have focused on obtaining
theoretical predictions in a rapidly rotating context (\cite{Lee1990},
\cite{Clement1994}, \cite{Townsend1997}) but more work is needed before these
effects are incorporated into mode identification tools such as FAMIAS
\cite{Zima2008}.

\begin{figure}[htbp]
\includegraphics[width=\columnwidth]{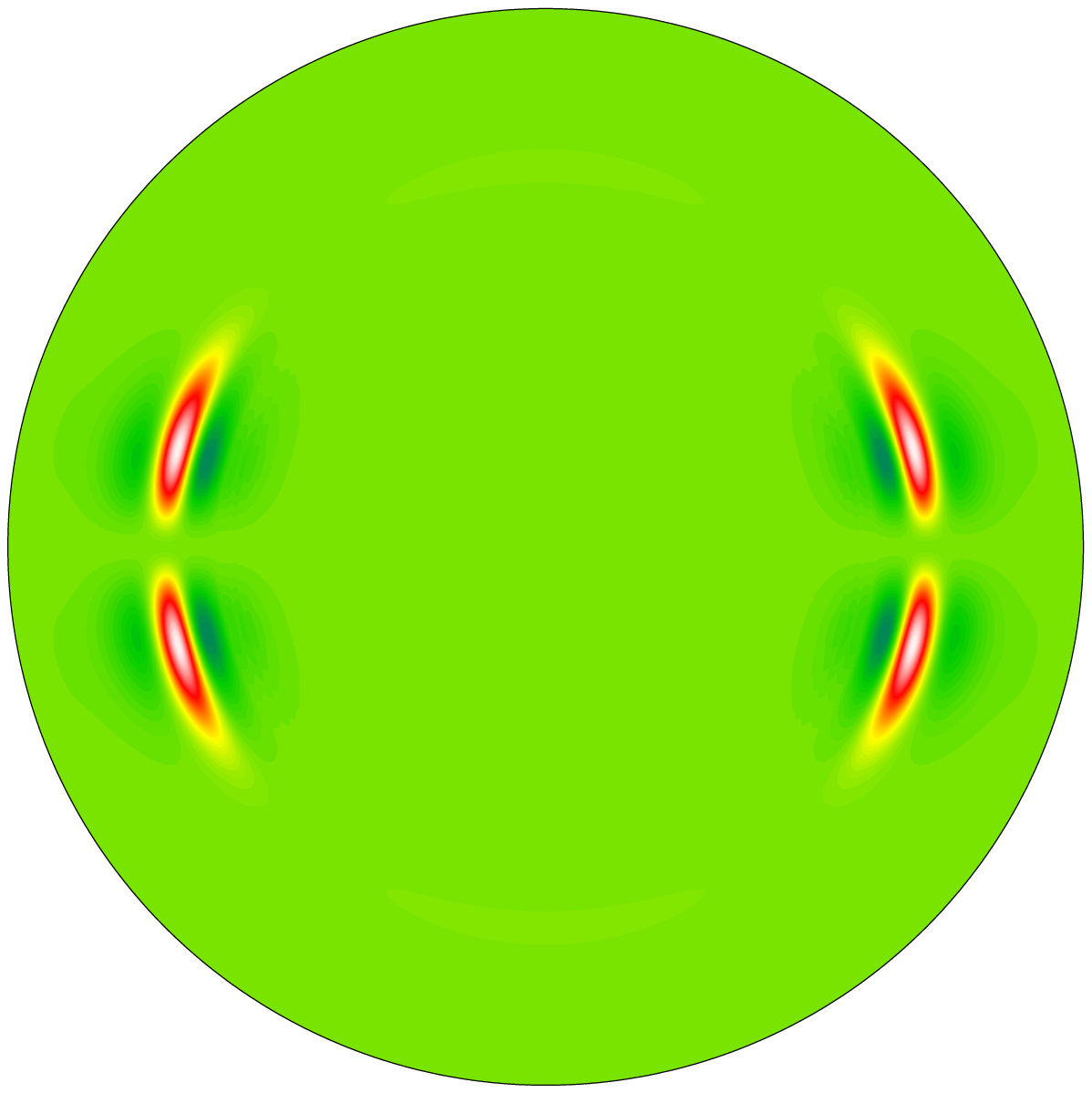} \\
\includegraphics[width=\columnwidth]{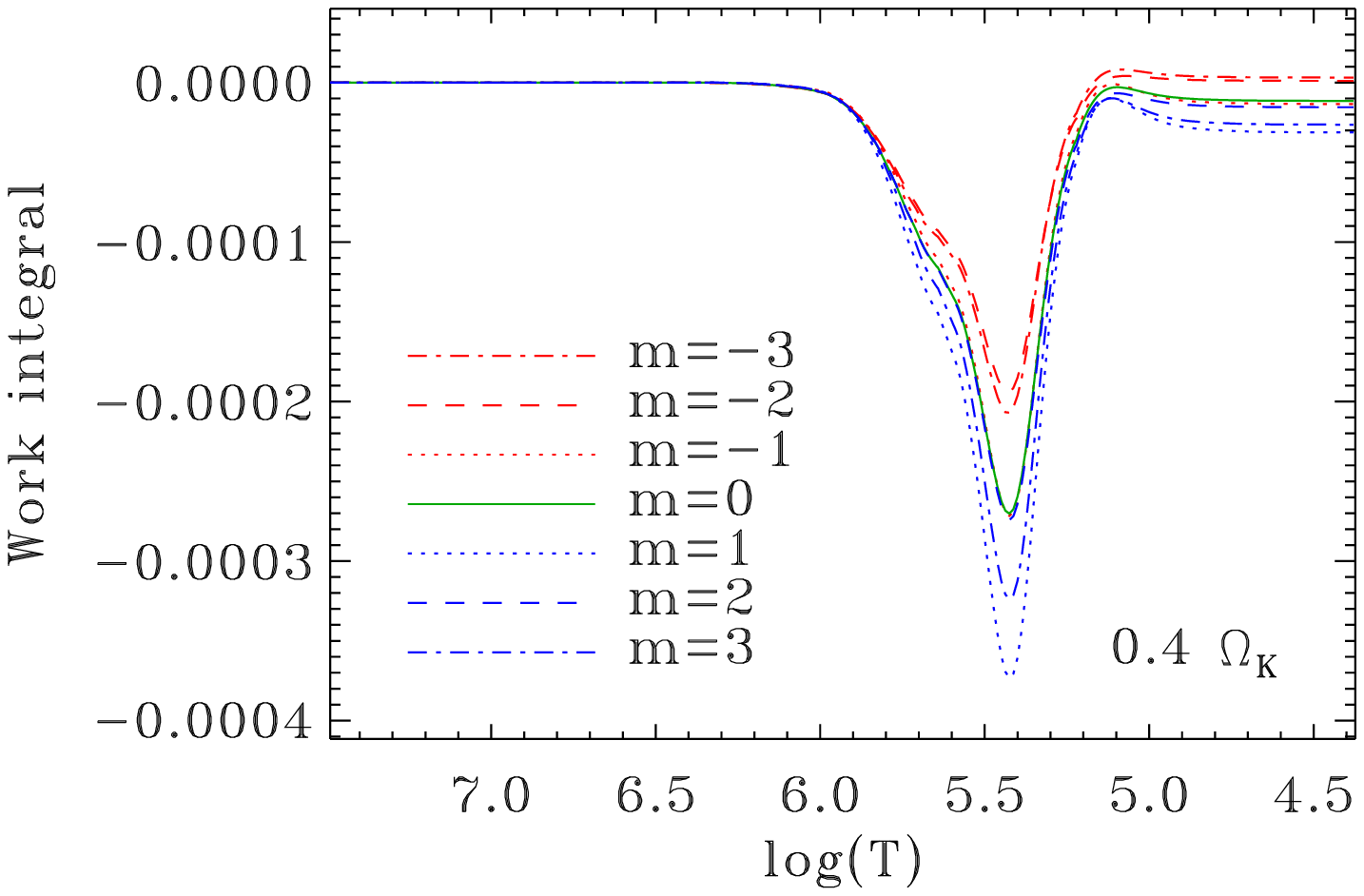} \\
\caption{\textit{(Upper panel)} Two dimensional map of regions which excite
(red/white) or damp (blue/black) a given island mode in an ESTER model.  The
radial coordinate in this plot is $\log T$ which stretches out the near surface
regions where most of the excitation and damping takes places.  \textit{(Lower
panel)}  Work integrals for a multiplet of modes at $\Omega = 0.4
\,\Omega_{\mathrm{K}}$.  As can be seen, some of the members of this multiplet
are stabilised by rotation. \label{fig:non_adiabatic}}
\end{figure}

An important ingredient in predicting theoretical photometric and spectroscopic
signatures of pulsation modes is the variations of effective temperature. 
However, this quantity can only be calculated through non-adiabatic pulsation
calculations.  Currently, the only studies that include these effects in rapidly
rotating stars are those based on \cite{Lee1995}, which use models based on a
Chandrasekhar expansion and focus on low frequency modes, and ongoing work which
involves using the TOP pulsation code (\cite{Reese2006}, \cite{Reese2009}) with
models from the ESTER code, an approach which is applicable to all modes. 
Figure~\ref{fig:non_adiabatic} shows a 2D excitation map as well as some 1D work
integrals of stable and unstable acoustic island-modes based on these latest
calculations. An added benefit of non-adiabatic calculations is knowing which
modes are unstable, which could provide additional constraints on the mode
identification. 

\section{Conclusion}

Rapid rotation plays a major role in the structure, evolution, and pulsations of
massive and intermediate mass stars.  Given that these stars are important for
many domains of astrophysics, it is essential to understand the impact of rapid
rotation.  Recent theoretical developments are leading to more and more
realistic models and a better understanding of their pulsation modes. 
Nonetheless, many questions remain unanswered, especially in the light of new
and spectacular observations.  Progress has been made in interpreting the
general characteristics of observed pulsation spectra, but more work is needed
in order to fully exploit the excellent data from current (MOST, CoRoT, Kepler,
Brite) and future missions (TESS, PLATO).

\begin{acknowledgement}
DRR is funded by the European Community's Seventh Framework Programme
(FP7/2007-2013) under grant agreement no. 312844 (SPACEINN), which is gratefully
acknowledged. Funding for the Stellar Astrophysics Centre is provided by The
Danish National Research Foundation (grant agreement No.: DNRF106).
\end{acknowledgement}

\bibliographystyle{epj}
\bibliography{Reese_D}

\begin{thebibliography}{80}

\bibitem{Royer2009}
F.~{Royer}, \emph{{On the Rotation of A-Type Stars}}, in \emph{The Rotation of
  Sun and Stars}, edited by J.P. {Rozelot}, C.~{Neiner} (2009), Vol. 765 of
  \emph{Lecture Notes in Physics, Berlin Springer Verlag}, pp. 207--230

\bibitem{DomicianoDeSouza2003}
A.~{Domiciano de Souza}, P.~{Kervella}, S.~{Jankov}, L.~{Abe}, F.~{Vakili},
  E.~{di Folco}, F.~{Paresce}, A\&A \textbf{407}, L47 (2003)

\bibitem{Kervella2006}
P.~{Kervella}, A.~{Domiciano de Souza}, A\&A \textbf{453}, 1059 (2006)

\bibitem{DomicianoDeSouza2012}
A.~{Domiciano de Souza}, M.~{Hadjara}, F.~{Vakili}, P.~{Bendjoya},
  F.~{Millour}, L.~{Abe}, A.C. {Carciofi}, D.M. {Faes}, P.~{Kervella},
  S.~{Lagarde} et~al., A\&A \textbf{545}, A130 (2012)

\bibitem{Peterson2006}
D.M. {Peterson}, C.A. {Hummel}, T.A. {Pauls}, J.T. {Armstrong}, J.A. {Benson},
  G.C. {Gilbreath}, R.B. {Hindsley}, D.J. {Hutter}, K.J. {Johnston},
  D.~{Mozurkewich} et~al., ApJ \textbf{636}, 1087 (2006)

\bibitem{Monnier2007}
J.D. {Monnier}, M.~{Zhao}, E.~{Pedretti}, N.~{Thureau}, M.~{Ireland},
  P.~{Muirhead}, J.P. {Berger}, R.~{Millan-Gabet}, G.~{Van Belle}, T.~{ten
  Brummelaar} et~al., Science \textbf{317}, 342 (2007),
  \texttt{arXiv:0706.0867}

\bibitem{Che2011}
X.~{Che}, J.D. {Monnier}, M.~{Zhao}, E.~{Pedretti}, N.~{Thureau},
  A.~{M{\'e}rand}, T.~{ten Brummelaar}, H.~{McAlister}, S.T. {Ridgway},
  N.~{Turner} et~al., ApJ \textbf{732}, 68 (2011)

\bibitem{Suarez2002}
J.C. {Su{\'a}rez}, E.~{Michel}, F.~{P{\'e}rez Hern{\'a}ndez}, Y.~{Lebreton},
  Z.P. {Li}, L.~{Fox Machado}, A\&A \textbf{390}, 523 (2002)

\bibitem{Salmon2014}
S.J.A.J. {Salmon}, J.~{Montalb{\'a}n}, D.R. {Reese}, M.A. {Dupret},
  P.~{Eggenberger}, A\&A \textbf{569}, A18 (2014)

\bibitem{vonZeipel1924}
H.~{von Zeipel}, MNRAS \textbf{84}, 665 (1924)

\bibitem{Lucy1967}
L.B. {Lucy}, {Zeitschrift f{\"u}r Astrophysic} \textbf{65}, 89 (1967)

\bibitem{EspinosaLara2011}
F.~{Espinosa Lara}, M.~{Rieutord}, A\&A \textbf{533}, A43 (2011)

\bibitem{Rieutord2005}
M.~{Rieutord}, B.~{Dintrans}, F.~{Ligni{\`e}res}, T.~{Corbard}, B.~{Pichon},
  \emph{{The ESTER project}}, in \emph{SF2A-2005: Semaine de l'Astrophysique
  Francaise} (2005), pp. 759--762

\bibitem{Rieutord2009}
M.~{Rieutord}, F.~{Espinosa Lara}, Communications in Asteroseismology
  \textbf{158}, 99 (2009)

\bibitem{EspinosaLara2013}
F.~{Espinosa Lara}, M.~{Rieutord}, A\&A \textbf{552}, A35 (2013)

\bibitem{Meynet2000}
G.~{Meynet}, A.~{Maeder}, A\&A \textbf{361}, 101 (2000),
  \texttt{astro-ph/0006404}

\bibitem{Zahn1992}
J.P. {Zahn}, A\&A \textbf{265}, 115 (1992)

\bibitem{Meynet2005}
G.~{Meynet}, A.~{Maeder}, \emph{{Rotation and Mixing in Massive Stars:
  Principles and Uncertainties}}, in \emph{The Nature and Evolution of Disks
  Around Hot Stars}, edited by R.~{Ignace}, K.G. {Gayley} (2005), Vol. 337 of
  \emph{Astronomical Society of the Pacific Conference Series}, pp. 15--+,
  \texttt{astro-ph/0409722}

\bibitem{Hunter2008}
I.~{Hunter}, I.~{Brott}, D.J. {Lennon}, N.~{Langer}, P.L. {Dufton},
  C.~{Trundle}, S.J. {Smartt}, A.~{de Koter}, C.J. {Evans}, R.S.I. {Ryans}, ApJ
  \textbf{676}, L29 (2008)

\bibitem{Brott2011b}
I.~{Brott}, C.J. {Evans}, I.~{Hunter}, A.~{de Koter}, N.~{Langer}, P.L.
  {Dufton}, M.~{Cantiello}, C.~{Trundle}, D.J. {Lennon}, S.E. {de Mink} et~al.,
  A\&A \textbf{530}, A116 (2011)

\bibitem{Aerts2014}
C.~{Aerts}, G.~{Molenberghs}, M.G. {Kenward}, C.~{Neiner}, ApJ \textbf{781}, 88
  (2014)

\bibitem{EspinosaLara2007}
F.~{Espinosa Lara}, M.~{Rieutord}, A\&A \textbf{470}, 1013 (2007)

\bibitem{Maeder2008}
A.~{Maeder}, C.~{Georgy}, G.~{Meynet}, A\&A \textbf{479}, L37 (2008)

\bibitem{Wang}
J.~{Wang}, C.~{Liang}, M.S. {Miesch}, {Journal of Computational Physics}  ({in
  preparation})

\bibitem{Ledoux1951}
P.~{Ledoux}, ApJ \textbf{114}, 373 (1951)

\bibitem{Schou1998}
J.~{Schou}, H.M. {Antia}, S.~{Basu}, R.S. {Bogart}, R.I. {Bush}, S.M. {Chitre},
  J.~{Christensen-Dalsgaard}, M.P. {di Mauro}, W.A. {Dziembowski},
  A.~{Eff-Darwich} et~al., ApJ \textbf{505}, 390 (1998)

\bibitem{Deheuvels2012}
S.~{Deheuvels}, R.A. {Garc{\'{\i}}a}, W.J. {Chaplin}, S.~{Basu}, H.M. {Antia},
  T.~{Appourchaux}, O.~{Benomar}, G.R. {Davies}, Y.~{Elsworth}, L.~{Gizon}
  et~al., ApJ \textbf{756}, 19 (2012)

\bibitem{Kurtz2014}
D.W. {Kurtz}, H.~{Saio}, M.~{Takata}, H.~{Shibahashi}, S.J. {Murphy},
  T.~{Sekii}, MNRAS \textbf{444}, 102 (2014), {See also these proceedings}

\bibitem{Saio1981}
H.~{Saio}, ApJ \textbf{244}, 299 (1981)

\bibitem{Soufi1998}
F.~{Soufi}, M.J. {Goupil}, W.A. {Dziembowski}, A\&A \textbf{334}, 911 (1998)

\bibitem{Reese2006}
D.~{Reese}, F.~{Ligni{\`e}res}, M.~{Rieutord}, A\&A \textbf{455}, 621 (2006),
  \texttt{astro-ph/0605503}

\bibitem{Ballot2010}
J.~{Ballot}, F.~{Ligni{\`e}res}, D.R. {Reese}, M.~{Rieutord}, A\&A
  \textbf{518}, A30 (2010)

\bibitem{Papaloizou1978}
J.~{Papaloizou}, J.E. {Pringle}, MNRAS \textbf{182}, 423 (1978)

\bibitem{Lee2006}
U.~{Lee}, MNRAS \textbf{365}, 677 (2006)

\bibitem{Rieutord2000}
M.~{Rieutord}, B.~{Georgeot}, L.~{Valdettaro}, Physical Review Letters
  \textbf{85}, 4277 (2000)

\bibitem{Dintrans2000}
B.~{Dintrans}, M.~{Rieutord}, A\&A \textbf{354}, 86 (2000)

\bibitem{Mirouh}
G.~{Mirouh}, \emph{{Gravito-inertial waves in a differentially rotating
  spherical shell}}, in \emph{These proceedings}, edited by {{Ballot}, J. and
  {Garc{\'\i}a}, R. A.} (2014)

\bibitem{Tassoul1980}
M.~{Tassoul}, ApJS \textbf{43}, 469 (1980)

\bibitem{Berthomieu1978}
G.~{Berthomieu}, G.~{Gonczi}, P.~{Graff}, J.~{Provost}, A.~{Rocca}, A\&A
  \textbf{70}, 597 (1978)

\bibitem{Lee1987}
U.~{Lee}, H.~{Saio}, MNRAS \textbf{224}, 513 (1987)

\bibitem{Ballot2012}
J.~{Ballot}, F.~{Ligni{\`e}res}, V.~{Prat}, D.R. {Reese}, M.~{Rieutord},
  \emph{{2D Computations of g-modes in Fast Rotating Stars}}, in \emph{Progress
  in Solar/Stellar Physics with Helio- and Asteroseismology}, edited by
  H.~{Shibahashi}, M.~{Takata}, A.E. {Lynas-Gray} (2012), Vol. 462 of
  \emph{Astronomical Society of the Pacific Conference Series}, p. 389

\bibitem{Dintrans1999}
B.~{Dintrans}, M.~{Rieutord}, L.~{Valdettaro}, Journal of Fluid Mechanics
  \textbf{398}, 271 (1999)

\bibitem{Takata2013}
M.~{Takata}, H.~{Saio}, Publ. of the Astronomical Society of Japan \textbf{65},
  68 (2013)

\bibitem{Saio2014}
H.~{Saio}, M.~{Takata}, Publ. of the Astronomical Society of Japan \textbf{66},
  58 (2014)

\bibitem{Takata2014}
M.~{Takata}, Publ. of the Astronomical Society of Japan \textbf{66}, 80 (2014)

\bibitem{Takata}
M.~{Takata}, H.~{Saio}, \emph{{Rosette modes of oscillations in rotating stars
  as a new aspect of rotation-pulsation interaction}}, in \emph{These
  proceedings}, edited by {{Ballot}, J. and {Garc{\'\i}a}, R. A.} (2014)

\bibitem{Lignieres2008}
F.~{Ligni{\`e}res}, B.~{Georgeot}, PRE \textbf{78}, 016215 (2008),
  \texttt{0803.1737}

\bibitem{Lignieres2009}
F.~{Ligni{\`e}res}, B.~{Georgeot}, A\&A \textbf{500}, 1173 (2009),
  \texttt{0903.1768}

\bibitem{Reese2009}
D.R. {Reese}, K.B. {MacGregor}, S.~{Jackson}, A.~{Skumanich}, T.S. {Metcalfe},
  A\&A \textbf{506}, 189 (2009)

\bibitem{Pasek2011}
M.~{Pasek}, B.~{Georgeot}, F.~{Ligni{\`e}res}, D.R. {Reese}, Physical Review
  Letters \textbf{107}, 121101 (2011)

\bibitem{Pasek2012}
M.~{Pasek}, F.~{Ligni{\`e}res}, B.~{Georgeot}, D.R. {Reese}, A\&A \textbf{546},
  A11 (2012)

\bibitem{Reese2008}
D.~{Reese}, F.~{Ligni{\`e}res}, M.~{Rieutord}, A\&A \textbf{481}, 449 (2008),
  \texttt{0801.4630}

\bibitem{MacGregor2007}
K.B. {MacGregor}, S.~{Jackson}, A.~{Skumanich}, T.S. {Metcalfe}, ApJ
  \textbf{663}, 560 (2007), \texttt{arXiv:0704.1275}

\bibitem{Ouazzani2013}
R.M. {Ouazzani}, M.J. {Goupil}, M.A. {Dupret}, J.P. {Marques}, A\&A
  \textbf{554}, A80 (2013)

\bibitem{Walker2005b}
G.A.H. {Walker}, R.~{Kuschnig}, J.M. {Matthews}, C.~{Cameron}, H.~{Saio},
  U.~{Lee}, E.~{Kambe}, S.~{Masuda}, D.B. {Guenther}, A.F.J. {Moffat} et~al.,
  ApJ \textbf{635}, L77 (2005)

\bibitem{Saio2007}
H.~{Saio}, C.~{Cameron}, R.~{Kuschnig}, G.A.H. {Walker}, J.M. {Matthews}, J.F.
  {Rowe}, U.~{Lee}, D.~{Huber}, W.W. {Weiss}, D.B. {Guenther} et~al., ApJ
  \textbf{654}, 544 (2007)

\bibitem{Cameron2008}
C.~{Cameron}, H.~{Saio}, R.~{Kuschnig}, G.A.H. {Walker}, J.M. {Matthews}, D.B.
  {Guenther}, A.F.J. {Moffat}, S.M. {Rucinski}, D.~{Sasselov}, W.W. {Weiss},
  ApJ \textbf{685}, 489 (2008)

\bibitem{Semaan2013}
T.~{Semaan}, J.~{Guti{\'e}rrez-Soto}, Y.~{Fr{\'e}mat}, A.M. {Hubert},
  C.~{Martayan}, J.~{Zorec}, \emph{{A Pulsational Study of a Sample of CoRoT
  Faint Be Stars}}, in \emph{Stellar Pulsations: Impact of New Instrumentation
  and New Insights}, edited by J.C. {Su{\'a}rez}, R.~{Garrido}, L.A. {Balona},
  J.~{Christensen-Dalsgaard} (2013), Vol.~31 of \emph{Advances in Solid State
  Physics}, p. 261

\bibitem{Breger2009}
M.~{Breger}, P.~{Lenz}, A.A. {Pamyatnykh}, MNRAS \textbf{396}, 291 (2009)

\bibitem{GarciaHernandez2009}
A.~{Garc{\'{\i}}a Hern{\'a}ndez}, A.~{Moya}, E.~{Michel}, R.~{Garrido}, J.C.
  {Su{\'a}rez}, E.~{Rodr{\'{\i}}guez}, P.J. {Amado}, S.~{Mart{\'{\i}}n-Ruiz},
  A.~{Rolland}, E.~{Poretti} et~al., A\&A \textbf{506}, 79 (2009)

\bibitem{Mantegazza2012}
L.~{Mantegazza}, E.~{Poretti}, E.~{Michel}, M.~{Rainer}, F.~{Baudin},
  A.~{Garc{\'{\i}}a Hern{\'a}ndez}, T.~{Semaan}, M.~{Alvarez}, P.J. {Amado},
  R.~{Garrido} et~al., A\&A \textbf{542}, A24 (2012)

\bibitem{GarciaHernandez2013}
A.~{Garc{\'{\i}}a Hern{\'a}ndez}, A.~{Moya}, E.~{Michel}, J.C. {Su{\'a}rez},
  E.~{Poretti}, S.~{Mart{\'{\i}}n-Ru{\'{\i}}z}, P.J. {Amado}, R.~{Garrido},
  E.~{Rodr{\'{\i}}guez}, M.~{Rainer} et~al., A\&A \textbf{559}, A63 (2013)

\bibitem{Suarez2014}
J.C. {Su{\'a}rez}, A.~{Garc{\'{\i}}a Hern{\'a}ndez}, A.~{Moya}, C.~{Rodrigo},
  E.~{Solano}, R.~{Garrido}, J.R. {Rod{\'o}n}, A\&A \textbf{563}, A7 (2014)

\bibitem{GarciaHernandez}
A.~{Garc{\'\i}a Hern{\'a}ndez}, F.~{Ligni{\`e}res}, L.~{Balona}, D.R. {Reese},
  M.J.P.F.G. {Monteiro}, J.C. {Su{\'a}rez}, J.~{Ballot}, \emph{{Patterns, an
  efficient way to analyse the p-mode content in rapidly rotating stars}}, in
  \emph{These proceedings}, edited by {{Ballot}, J. and {Garc{\'\i}a}, R. A.}
  (2014)

\bibitem{Lignieres2010}
F.~{Ligni{\`e}res}, B.~{Georgeot}, J.~{Ballot}, Astronomische Nachrichten
  \textbf{331}, 1053 (2010)

\bibitem{Reese}
D.R. {Reese}, F.~{Ligni{\`e}res}, J.~{Ballot}, M.A. {Dupret}, C.~{Barban},
  C.~{van~'t Veer-Menneret}, K.B. {MacGregor}, A\&A  ({submitted})

\bibitem{Deupree2012}
R.G. {Deupree}, D.~{Casta{\~n}eda}, F.~{Pe{\~n}a}, C.I. {Short}, ApJ
  \textbf{753}, 20 (2012)

\bibitem{Bedding}
T.~{Bedding}, S.~{Murphy}, D.~{Stello}, \emph{{Unlocking the secrets of gamma
  Doradus and delta Scuti stars using echelle diagrams}}, in \emph{These
  proceedings}, edited by {{Ballot}, J. and {Garc{\'\i}a}, R. A.} (2014)

\bibitem{Reese2009b}
D.R. {Reese}, M.J. {Thompson}, K.B. {MacGregor}, S.~{Jackson}, A.~{Skumanich},
  T.S. {Metcalfe}, A\&A \textbf{506}, 183 (2009)

\bibitem{Townsend2003}
R.H.D. {Townsend}, MNRAS \textbf{343}, 125 (2003)

\bibitem{Daszynska_Daszkiewicz2007}
J.~{Daszynska-Daszkiewicz}, W.A. {Dziembowski}, A.A. {Pamyatnykh}, Acta
  Astronomica \textbf{57}, 11 (2007)

\bibitem{Reese2013}
D.R. {Reese}, V.~{Prat}, C.~{Barban}, C.~{van~'t Veer-Menneret}, K.B.
  {MacGregor}, A\&A \textbf{550}, A77 (2013)

\bibitem{Telting1998}
J.H. {Telting}, C.~{Schrijvers}, A\&A \textbf{339}, 150 (1998)

\bibitem{Poretti2009}
E.~{Poretti}, E.~{Michel}, R.~{Garrido}, L.~{Lef{\`e}vre}, L.~{Mantegazza},
  M.~{Rainer}, E.~{Rodr{\'{\i}}guez}, K.~{Uytterhoeven}, P.J. {Amado},
  S.~{Mart{\'{\i}}n-Ruiz} et~al., A\&A \textbf{506}, 85 (2009)

\bibitem{Themessl}
N.~{Theme{\ss}l}, \emph{{Frequency analysis and spectroscopic mode
  identification of the delta Scuti star 4 CVn}}, in \emph{These proceedings},
  edited by {{Ballot}, J. and {Garc{\'\i}a}, R. A.} (2014)

\bibitem{Lee1990}
U.~{Lee}, H.~{Saio}, ApJ \textbf{349}, 570 (1990)

\bibitem{Clement1994}
M.J. {Clement}, \emph{{Pulsation in Rapidly Rotating Stars}}, in \emph{IAU
  Symp. 162: Pulsation; Rotation; and Mass Loss in Early-Type Stars}, edited by
  L.A. {Balona}, H.F. {Henrichs}, J.M. {Le Contel} (1994), pp. 117--+

\bibitem{Townsend1997}
R.H.D. {Townsend}, MNRAS \textbf{284}, 839 (1997)

\bibitem{Zima2008}
W.~{Zima}, Communications in Asteroseismology \textbf{155}, 17 (2008)

\bibitem{Lee1995}
U.~{Lee}, I.~{Baraffe}, A\&A \textbf{301}, 419 (1995)

\end{thebibliography}

\end{document}